\documentclass[12pt]{iopart}
\usepackage{amssymb}
\usepackage[amssymb]{SIunits}
\usepackage{graphicx}
\usepackage{xspace}
\usepackage{bm}
\usepackage[colorlinks,urlcolor=blue]{hyperref}
\usepackage[all]{hypcap}
\newcommand*{\doi}[1]{\href{http://dx.doi.org/\detokenize{#1}}{\detokenize{#1}}}

\begin{document}

\title{Comparison of plasma response models for RMP effects on the divertor and scrape-off layer in KSTAR}

\author{H. Frerichs${}^1$, J. Van Blarcum${}^2$, T. Cote${}^3$, S.K. Kim${}^4$, Y.Q. Liu${}^3$, S.M. Yang${}^4$}

\address{${}^1$ Department of Nuclear Engineering \& Engineering Physics, University of Wisconsin - Madison, Madison, WI, USA}
\address{${}^2$ ITER Organization, St. Paul Lez Durance Cedex, France}
\address{${}^3$ General Atomics, San Diego, CA, USA}
\address{${}^4$ Princeton Plasma Physics Laboratory, Princeton, NJ, USA}

\ead{hfrerichs@wisc.edu}

\begin{abstract}
Resonant magnetic perturbations (RMPs) are beneficial for control of edge localized modes (ELMs) in tokamaks.
Nevertheless, a side effect is the appearance of a helical striations in the particle and heat loads onto divertor targets.
The extent and field line connection of these striations is significantly altered by the plasma response to external perturbations.
For an ELM suppressed H-mode plasma at KSTAR, magnetic footprints are computed by FLARE based on plasma response from GPEC, MARS-F, M3D-C1 and JOREK with substantial differences in the resulting footprints (from 2 cm to 14 cm).
This is reflected in EMC3-EIRENE simulations of the resulting heat loads: it is found that either the peak value or the extent of the striations appear to be overestimated compared to IRTV measurements.
Reasonable agreement can only be achieved for the smallest footprint for lower input power and lower cross-field transport, or for higher upstream density and radiative power losses.
\end{abstract}

\vspace{2pc}
\noindent{\it Keywords}: resonant magnetic perturbations, plasma response, divertor heat loads

\def\vec#1{\ensuremath{{\bf #1}}\xspace}
\def\Poincare{Poincar\'e\xspace}
\def\psiN{\ensuremath{\psi_{N}}\xspace}
\def\minPsiN{\ensuremath{\mathrm{min}_F \, \psi_{N}}\xspace}
\def\minR{\ensuremath{\mathrm{min} \, \mathcal{R}}\xspace}
\def\Lpt{\ensuremath{L_{\mathrm{pt}}}\xspace}
\def\nsepx{\ensuremath{n_{\mathrm{sepx}}}\xspace}
\def\PSOL{\ensuremath{P_{\mathrm{SOL}}}\xspace}
\def\Pheat{\ensuremath{P_{\mathrm{heat}}}\xspace}
\def\qmax#1{\ensuremath{q_{\textnormal{max}}^{\textnormal{(#1)}}}\xspace}
\def\csput{\ensuremath{c_{\mathrm{sput}}}\xspace}
\def\qboundary{\ensuremath{q_{\mathrm{a}}}\xspace}

\section{Introduction}

The high confinement (H-mode) regime is the baseline for burning plasma operation in ITER \cite{ITR-18-003, ITR-24-005}.
However, a major challenge associated with H-mode operation are edge localised modes (ELMs) which require a control scheme \cite{Evans2013a, Lang2013, Loarte2014} in order to avoid damage to plasma facing components due to rapid energy loss from the pedestal region.
One particular control scheme which has been successfully reproduced in a number of present tokamaks is the application of resonant magnetic perturbations (RMPs) \cite{Kirk2013, Evans2015}.
The scrape-off layer (SOL) and divertor play a major role for the path to tokamak burning plasma operation \cite{Krieger2025}, but a side effect of RMP application is the appearance of helical striation patterns in the particle and heat loads onto divertor targets.
These have been observed early on at DIII-D \cite{Evans2008}, and simulations for ITER have shown that they can affect the efficiency of power dissipation \cite{Frerichs2020, Frerichs2021a, Frerichs2024}.

Accurate predictions of divertor heat loads must include the plasma response to the external perturbations.
The plasma response tends to reduce the radial excursion of perturbed field lines due to screening effects, but it can also result in field amplification \cite{Liu2010, Haskey2014, Ryan2015}.
In particular, field amplification near the separatrix can have a significant impact on the field line connection to the divertor targets.
Over the years, a number of different plasma response models (linear vs. non-linear, ideal vs. resistive MHD, single fluid vs. two fluid ...) have been developed and compared to each other \cite{Turnbull2012, Turnbull2013, Reiman2015}, however, mostly with a focus on the plasma core or edge / pedestal top.
Implications of different plasma response models for the SOL and divertor, on the other hand, have only been briefly explored for DIII-D \cite{Lore2017}.
So far, however, with inconclusive results with respect to reproducing upstream and downstream observations at the same time.

In the present study we compare different plasma response models with respect to their impact on divertor footprints for a KSTAR H-mode plasma with ELM suppression by RMPs.
In section \ref{sec:magnetic_geometry}, magnetic footprints (field line tracing) are constructed with FLARE \cite{Frerichs2024a} based on plasma response from GPEC \cite{Park2017, Park2018a}, MARS-F \cite{Liu2000, Liu2010}, M3D-C1 \cite{Jardin2008, Ferraro2012} and JOREK \cite{Czarny2008, Hoelzl2021}.
All of these plasma response models employ full toroidal geometry.
GPEC and MARS-F are linear plasma response models, while both M3D-C1 and JOREK are run in linear mode for this comparison (or a linear-like mode, see below).
Inputs are equilibrium profiles (density, temperature, current) and 3D coil currents for magnetic field perturbations.
As extension to ideal perturbed equilibrium code (IPEC) \cite{Park2007}, GPEC calculates the kinetic force balance including self-consistent neoclassical toroidal viscosity (NTV) torque.
For MARS-F, we choose the single-fluid resistive-rotating plasma (RRP) assumption where the Spitzer resistivity model is adopted and the experimentally measured toroidal rotation of the plasma is included in the modeling.
Although M3D-C1 can solve the non-linear visco-resistive two-fluid MHD equations, a non-linear M3D-C1 free-boundary equilibrium calculation for this configuration failed to converge.
This has been attributed to the near double null shape.
Instead, we use a linear, single-fluid fixed boundary M3D-C1 calculation.
Non-linear extended MHD equations are implemented in JOREK, but here we force a linear-like result by applying RMPs with $2.5 \, \%$ strength and scaling the result by 40.
Then, EMC3-EIRENE \cite{Feng2004, Feng2014} simulations are conducted in order to evaluate the resulting heat loads onto divertor targets.
The simulation setup is described in section \ref{sec:setup}, and simulation results are discussed in section \ref{sec:results}.


\section{Magnetic geometry} \label{sec:magnetic_geometry}

The foundation of the present analysis is KSTAR discharge 30306 with a kinetic EFIT equilibrium reconstruction at 7850 ms.
The plasma current is $I_p = 0.51 \, \mega\ampere$ with a magnetic field of $1.79 \, \tesla$ on axis and a safety factor of $q_{95} = 5.24$.
The configuration is biased towards the lower X-point with a separation of $2 \, \centi\meter$ between the primary and secondary separatrix at the outboard midplane.
The outer strike point is located on the diagonal divertor target (carbon), which is monitored by an infrared camera \cite{Lee2017}.
RMPs are applied with toroidal mode number $n = 1$ with a $90 \, \deg$ phase shift between upper, middle and lower row of coils.
Coil currents are $2.67 \, \kilo\ampere$ in the upper and lower row, and $2 \, \kilo\ampere$ in the middle row. Each coil has 2 turns.

The magnetic geometry sets the stage for the boundary plasma.
Important aspects are the length of a field line until it is intercepted by a divertor target and where it strikes.
Beyond the last closed flux surface (LCFS), field lines are diverted to dedicated targets.
This happens within one poloidal turn in traditional single- (or double-) null configurations without RMPs.
Once RMPs are present, however, magnetic island chains appear on resonant surfaces (with integer numbers for toroidal and poloidal turns of a field line), and overlap between neighboring island chains results in regions with chaotic field line behavior.
Nevertheless, there is order in chaos: of particular intersect is a class of field lines that asymptotically approach the X-point either in forward or backward direction.
This is a generalization of what is known as the magnetic separatrix in tokamak configurations without RMPs.
These field lines form a manifold that other field lines cannot cross.
Thus, it separates field lines on the inside from the outside.

\begin{figure}
\begin{center}
\includegraphics[width=160mm]{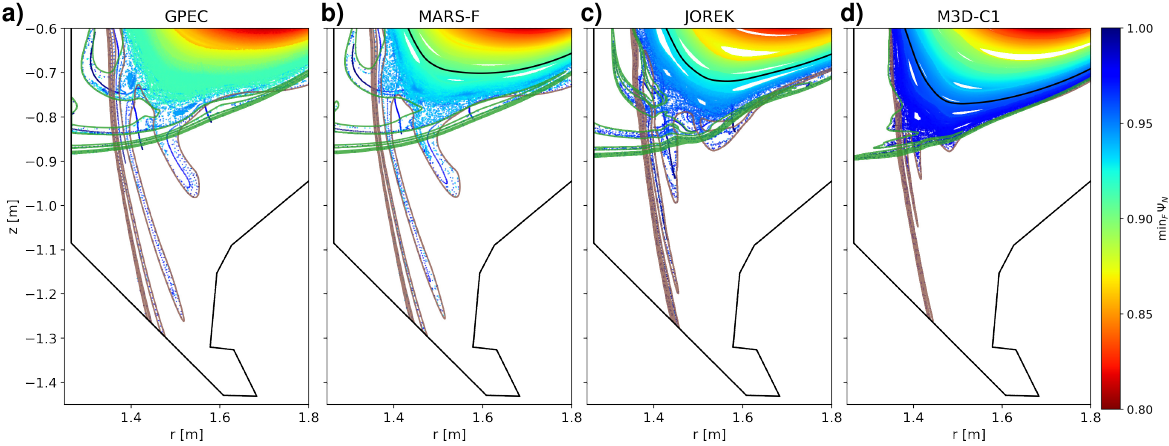}
\caption{Perturbed magnetic separatrix (dark green, brown) for different plasma response models. The puncture points of a high density \Poincare plot are colored based on the minimum of \psiN taken along the corresponding field line.
The core boundary for the EMC3-EIRENE simulations is shown in black.}
\label{fig:invariant_manifolds}
\end{center}
\end{figure}

Figure \ref{fig:invariant_manifolds} shows the forward (green) and backward (brown) branches of the magnetic separatrix for different plasma response models: GPEC, MARS-F, JOREK and M3D-C1.
It can be seen that both GPEC and MARS-F produce lobes that extend far into the scrape-off layer (SOL).
JOREK and M3D-C1, on the other hand, produce smaller lobes for the same external perturbation. 
Figure \ref{fig:invariant_manifolds} also shows high density \Poincare plots.
These are generated by launching 1024 field lines between \psiN = 0.8 and 1 at the outboard midplane, and then coloring all puncture points according to $\mathcal{R} \, = \, \minPsiN$, i.e. the minimum of \psiN (the normalized equilibrium poloidal flux) taken along the corresponding field line.
It can be seen how field lines are guided by the perturbed separatrix from the plasma edge towards the divertor targets on the first wall.
These {\it open} field lines form a new helical scrape-off layer and they are different from traditional SOL field lines in that they take one poloidal turn to connect into the plasma ($\psiN < 1$) and at least one additional poloidal turn before the leave and connect to (one of) the other target(s), i.e. the connection length is $\Lpt \ge 2$.

\begin{figure}
\begin{center}
\includegraphics[width=160mm]{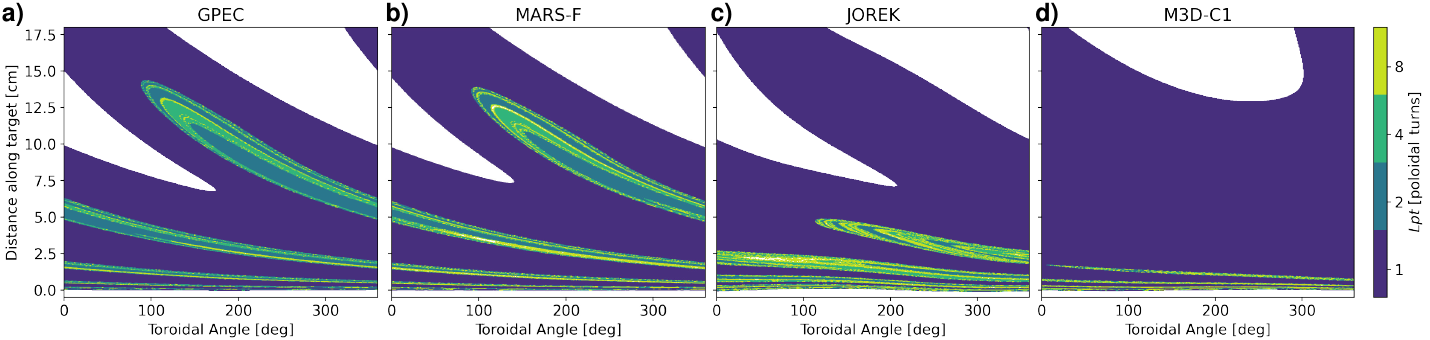}
\caption{Magnetic footprints on the outer lower divertor target. The distance along the target is measured from the strike point of the reference configuration without RMPs. The traditional SOL with one poloidal turn manifests as an extension of the new helical SOL. The far SOL connects with 1/2 poloidal turn from the lower outer target to the upper outer target.}
\label{fig:magnetic_footprints}
\end{center}
\end{figure}

The resulting footprints on the outer lower divertor target are shown in figure \ref{fig:magnetic_footprints} for the 4 plasma response models.
The new helical SOL is surrounded by traditional SOL (purple) with field lines outside the separatrix which connect the outer and inner lower divertor targets within one poloidal turn.
White colors indicate a short connection ($\Lpt \lesssim 1/2$) between the lower and upper outer divertor targets in disconnected double null configuration.
In the following, we define the footprint size $S$ as the maximum distance along the target where $\Lpt \ge 2$.
As suggested by the large lobes for GPEC and MARS-F in figure \ref{fig:invariant_manifolds}, these models produce the largest footprints: $S \, = \, 14.4 \, \centi\meter$ and $13.9 \, \centi\meter$, respectively.
The similarity of these two predictions is remarkable, in particular considering the sensitivity with respect to equilibrium truncation in GPEC required for the internal representation in magnetic coordinates as reported earlier \cite{Frerichs2023}.
Also, smoothing of the plasma boundary around the X-point is required for MARS-F, and this introduces a similar sensitivity there (see \ref{sec:marsf_boundary_scan}).
For JOREK, on the other hand, we find a much smaller footprint of $S = 4.9 \, \centi\meter$ while the striation pattern is barely visible for M3D-C1 with $S = 1.7 \, \centi\meter$.

\begin{table}
\footnotesize
\begin{center}\begin{tabular}{l|lll}
Model        & $S \, [\centi\meter]$  & $S_1 \, [\centi\meter]$  & \minR \\
\hline \\
Biot-Savart  & 16.5                   & 16.1  & 0.515 \\
GPEC         & 14.4                   & 10.5  & 0.934 \\
MARS-F       & 13.9                   & 9.8   & 0.926 \\
JOREK        &  4.9                   & 5.7   & 0.939 \\
M3D-C1       &  1.7                   & 1.0   & 0.968 \\
\end{tabular}\end{center}
\caption{Footprint parameters for different plasma response models: footprint size (S) and its approximation $S_1 \, = \, s(1 + \max M)$ based on the Melnikov integral $M(\varphi_0)$, as well as the deepest radial excursion from the target (\minR).
The Melnikov integral represents the accumulated displacement $\Delta \psiN$ along an unperturbed field line on the separatrix and depends on the toroidal angle $\varphi_0$ of a reference point (e.g. at the outboard midplane).
The footprint size is then approximated by taking the maximum over $\varphi_0$ and mapping the corresponding radial coordinate $\psiN = 1 \, + \, \max M$ onto the outer divertor target as distance $s(\psiN)$ from the equilibrium strike point.
}
\label{tab:footprint_parameters}
\end{table}

\begin{figure}
\begin{center}
\includegraphics[width=160mm]{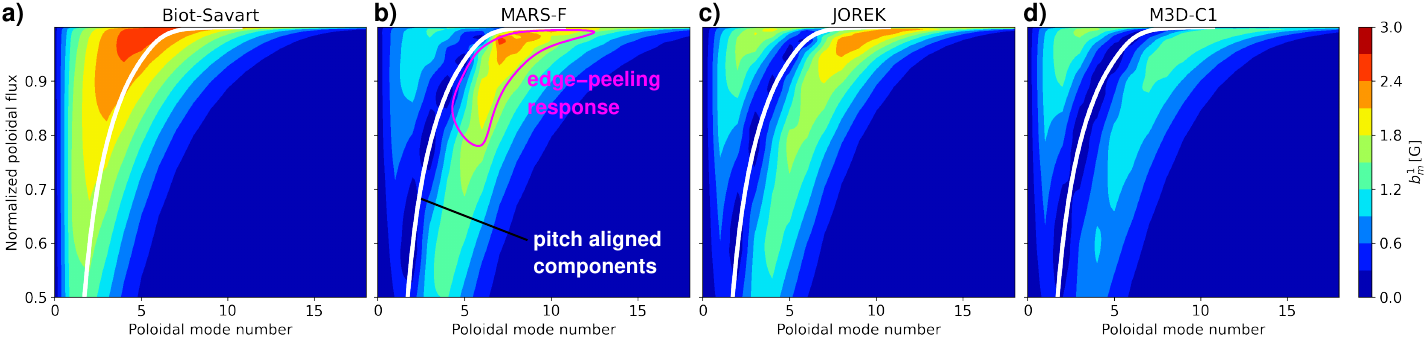}
\caption{Fourier harmonics of the external perturbation field (a) and the total perturbation field including plasma response from different models (b-d).}
\label{fig:fft}
\end{center}
\end{figure}

For smooth and sufficiently small perturbations, $S$ can be estimated from the Melnikov integral \cite{Wiggins2003, Joseph2008, Cahyna2010} which is related to the pitch aligned component in the limit $\psiN \rightarrow 1$.
This is a useful approximation for vacuum perturbation fields, but it can be inaccurate when screening close to the separatrix is included \cite{Cahyna2011}.
Nevertheless, it qualitatively explains the differences between the models as summarized in table \ref{tab:footprint_parameters}.
Figure \ref{fig:fft} shows the poloidal harmonics $b^1_m (\psiN)$ of the normal component of the total perturbation field evaluated along an equilibrium flux surface at \psiN.
It can be seen that the pitch aligned components are substantially screened in all plasma response models in comparison to the externally applied field.
As the resonant components determine the size of magnetic island chains and thus the potential for overlap, a rather narrow chaotic region with small radial field line excursion is found once plasma response is included (see \minR in table \ref{tab:footprint_parameters}).

While the field response inside the separatrix remains relatively similar across models, it is how the perturbation field extends beyond the plasma boundary that determines the magnetic footprint.
Qualitatively, we find that a larger footprint size $S$ is correlated with a stronger edge-peeling response (which in turn is correlated with ELM control).
In the JOREK simulation in this study, the magnetic flux perturbation is fixed on the computational boundary \cite{Becoulet2012, Orain2013}.
The computational boundary is guided by the disconnected double null layout of the equilibrium flux surfaces, and it is located at a distance of $4.7 \, \centi\meter$ from the separatrix at the outboard midplane.
This approach can constrain the plasma response and impact the magnetic lobe structure in the SOL.
For the M3D-C1 simulation, on the other hand, the computational boundary is a convex hull which is further away from the plasma.
Thus, the fixed boundary condition imposes less of a constraint on the plasma response around the separatrix in that case.
Recent studies using JOREK-STARWALL \cite{Hoelzl2012, Mitterauer2022} have demonstrated that a fully self-consistent free-boundary implementation, which couples the plasma to the vacuum region and to conducting structures, is important for reproducing experimental field responses in the SOL region.
The dedicated investigation of the impact of the fixed vs. free-boundary RMP constraints on footprint modeling will be important future work.
For now, we focus on comparing the resulting impact on divertor heat loads.




\section{Simulation setup} \label{sec:setup}

\begin{figure}
\begin{center}
\includegraphics[width=75mm]{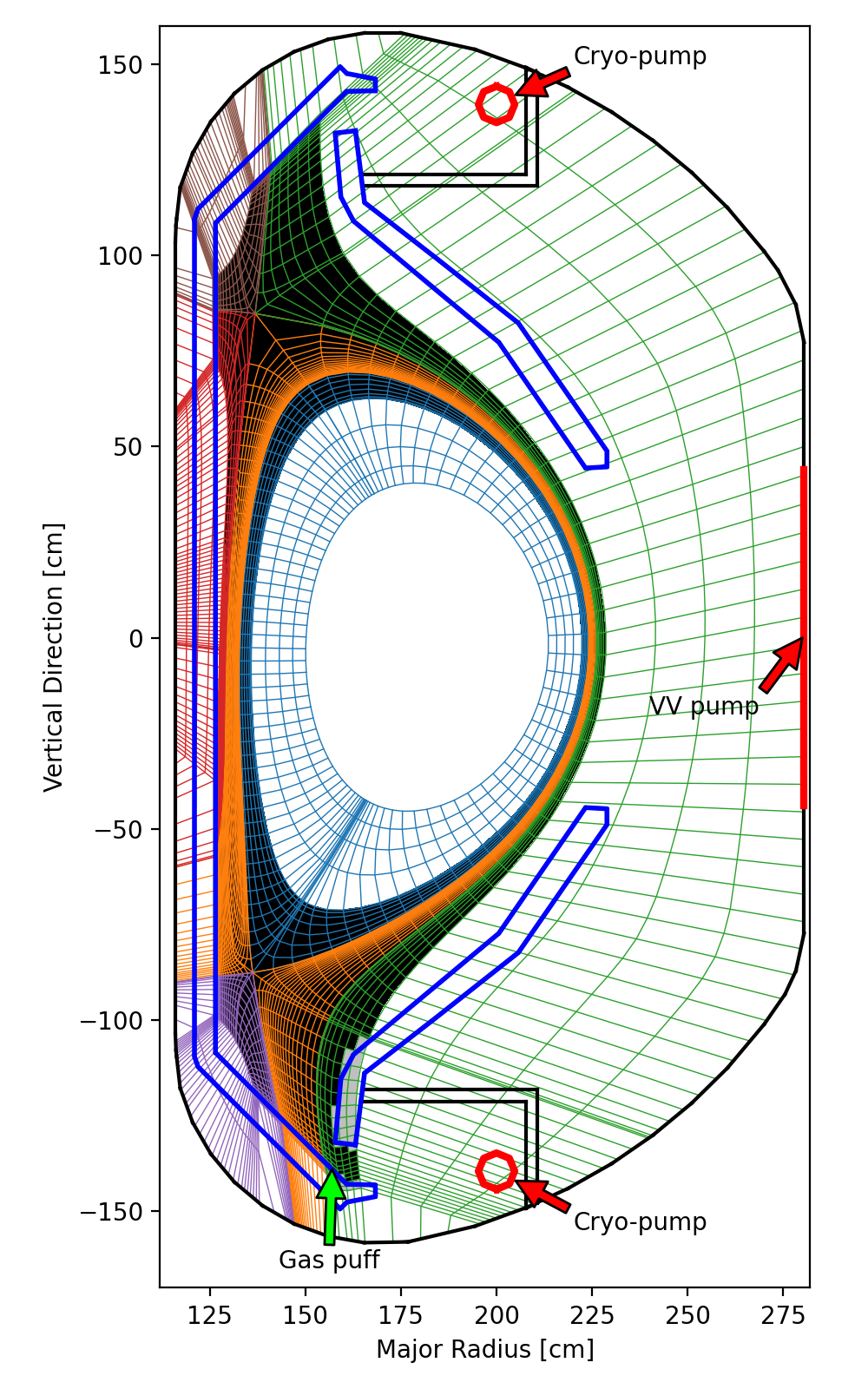}
\caption{Cross-section of computational mesh (lower resolution for visualization purposes) with block-structured layout in double null configuration. Plasma cells (EMC3) are shown in black. The mesh for neutral particles (EIRENE) is extended into the core and to the vacuum vessel.}
\label{fig:mmesh_rzslice}
\end{center}
\end{figure}

Before plasma boundary simulations with EMC3-EIRENE can be conduced, a magnetic mesh needs to be generated for fast field line reconstruction.
First, a pair of good flux surfaces for the core-edge interface is determined from \Poincare plots, and a block-structured base mesh with disconnected double null layout is constructed (see figure \ref{fig:mmesh_rzslice}).
The blocks are generated from the primary and secondary separatrix of the reference configuration without RMPs.
The simulation domain is split into 9 toroidal blocks $40 \, \deg$ in which flux tubes are then constructed by tracing perturbed field lines from each base mesh.
The 3-D mesh is extended into the core in order to account for charge exchange and ionization of fast neutral particles, and up to the vacuum vessel in order to account for exhaust of neutral particles through cryo-pumps.
The absorption probabilities are set to $2.06 \, \%$ for the two cryo-pumps which corresponds to a pumping speed of $25 \, \meter^3 \, \second^{-1}$.
The pumping surface in the middle mimics a vacuum vessel pump with $13.9 \, \meter^3 \, \second^{-1}$ for which the corresponding absorption probability is $0.29 \, \%$.
This is the same as in SOLPS simulations for KSTAR \cite{Hwang2020}.

Classical transport along field lines is assumed without flux limiters at this point.
Anomalous cross-field transport is modeled with a diffusive ansatz with model parameters $D = 0.3 \, \meter^2 \, \second^{-1}$ for particles and $\chi = 1.0 \, \meter^2 \, \second^{-1}$ for energy.
Deuterium is assumed to be completely recycled from the wall surfaces.
Carbon is sputtered with a yield of $\csput = 2 \, \%$ and is assumed to be completely adhered once it comes back to the wall surfaces.
The plasma density at the former separatrix (approximated from the last cell layer in the core block) is controlled via feedback adjustment of the gas puff strength at the outer target.
The total power into the SOL is set to $\PSOL = 3 \, \mega\watt$, which assumes $30 \, \%$ of the heating power $\Pheat = 4.3 \, \mega\watt$ has already been radiated in the core.
The power is then equally distributed between electrons and ions.
The particle flux from the core remains fixed at $3.4 \cdot 10^{20} \, \second^{-1}$ which corresponds to the particle source from neutral beam injections. 


\section{Heat loads} \label{sec:results}

\begin{figure}
\begin{center}
\includegraphics[width=160mm]{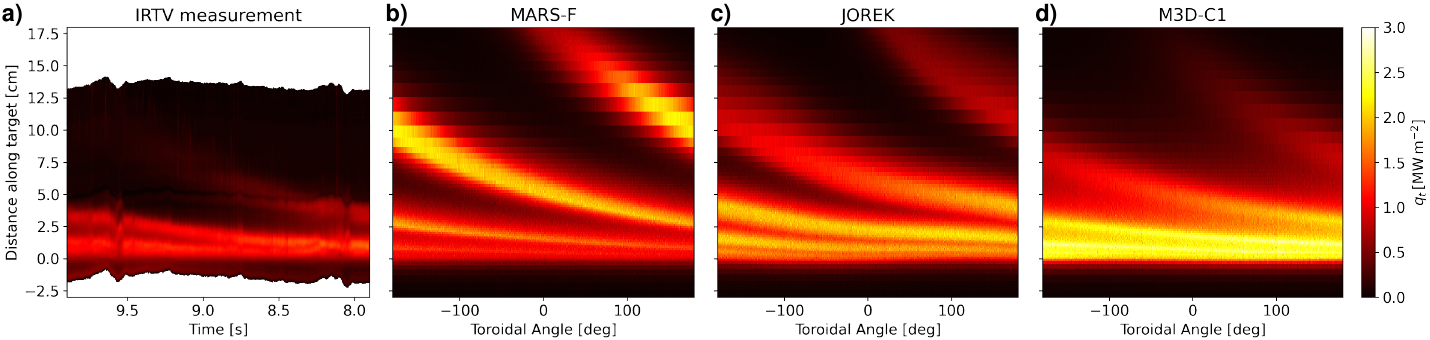}
\caption{(a) IRTV measurement of the heat load by slow rotation of the external perturbation with corrections for a drifting equilibrium \cite{VanBlarcum2025}. (b-d) Heat loads on the outer lower divertor target for $\nsepx = 0.5 \cdot 10^{19} \, \meter^{-3}$ and $\PSOL = 3 \, \mega\watt$ for different plasma response models.}
\label{fig:heat_load_benchmark_nsepx05_P30}
\end{center}
\end{figure}

Figure \ref{fig:heat_load_benchmark_nsepx05_P30} shows the resulting heat loads on the outer divertor target for a separatrix density of $\nsepx = 0.5 \cdot 10^{19} \, \meter^{-3}$ and the other simulation parameters described above.
No simulations have been conducted for the GPEC plasma response because of the similarities with the MARS-F magnetic geometry.
As expected from the magnetic footprints in figure \ref{fig:magnetic_footprints}, the heat load striation pattern with the MARS-F plasma response is larger than that with the JOREK or M3D-C1 plasma response.
However, the distinction between the new helical SOL ($\Lpt \ge 2$) and the traditional SOL ($\Lpt = 1$) is not as clear:
heat loads extend into the latter helically outwards from the tip of the lobe due to cross-field transport.
In comparison to the IRTV measurement on the left side of figure \ref{fig:heat_load_benchmark_nsepx05_P30}, it can be seen that 1) peak heat loads exceed the measurement by a factor of 2 for all plasma response models, and 2) the striation pattern is larger in all models with the M3D-C1 plasma response providing the closest match.
Given a magnetic footprint size of only $1.7 \, \centi\meter$, which would appear to be too small from a visual evaluation of figure \ref{fig:heat_load_benchmark_nsepx05_P30} (a), it is evident that cross-field transport into the traditional SOL competes with exhaust along the new helical SOL, and that estimating the footprint size from the heat load pattern is not straightforward.
The overestimation of the MARS-F footprint, on the other hand, is consistent with an overestimation of the computed plasma displacement near the separatrix along the outboard midplane in MARS-F when compared to thermal electron temperature perturbation measured by the electron-cyclotron emission imaging (ECEI) system \cite{Liu2024}.

\begin{figure}
\begin{center}
\includegraphics[width=160mm]{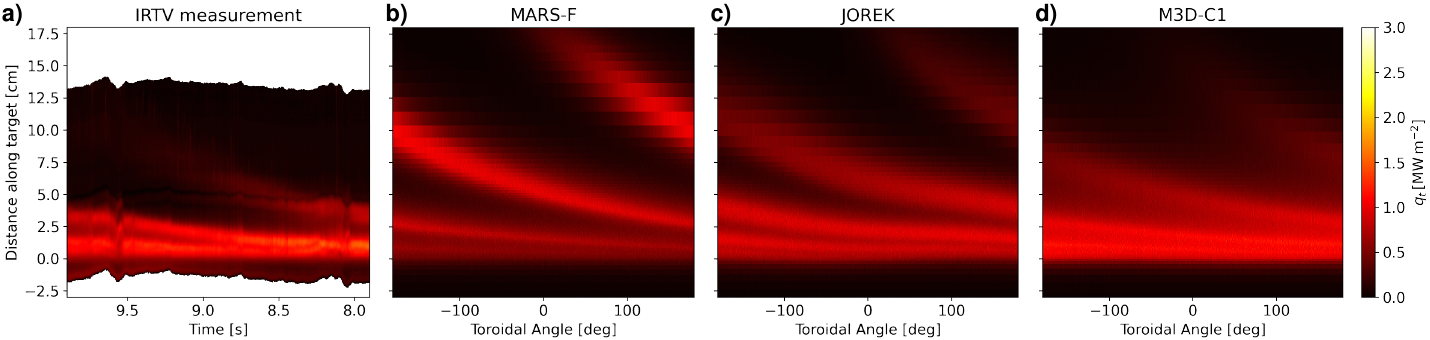}
\caption{(a) The same as figure \ref{fig:heat_load_benchmark_nsepx05_P30} (a). (b-d) Heat loads on the outer lower divertor target for $\nsepx = 0.5 \cdot 10^{19} \, \meter^{-3}$ and $\PSOL = 1.5 \, \mega\watt$ for different plasma response models.}
\label{fig:heat_load_benchmark_nsepx05_P15}
\end{center}
\end{figure}

Given the two key findings presented above, we will now explore possible reasons for misrepresentations in the simulations.
First, let us consider the power balance.
Out of the $3 \, \mega\watt$ power input to the SOL, nearly $90 \, \%$ is deposited on the wall surfaces while only $10 \, \%$ is radiated by recycled neutrals (the contributions from carbon are negligible) with similar numbers in all simulations.
This is consistent with an attached plasma at low density.
However, only $0.41 \, \mega\watt$ are observed by IRTV on the outer divertor target.
This could imply either a) a strong bias towards the inner target, b) an underestimation of the radiative power losses in the SOL in the simulations, or c) an overestimation of \PSOL (underestimation of radiative power losses in the core or NBI shine-through despite no significant losses being predicted by NUBEAM \cite{Pankin2004}).
Particle drifts ($\vec{E} \times \vec{B}$ and $\nabla B$) may contribute to a shift towards the inner target, but those a presently not available in EMC3-EIRENE.
On the other hand, intrinsic impurities may result in additional power losses beyond what is found in the simulations from $2 \, \%$ carbon sputtering.
This could be either due to a higher net sputtering rate for the wall material, other impurity species present in the device, or different temperature and electron density conditions which may produce more radiation.

\begin{figure}
\begin{center}
\includegraphics[width=160mm]{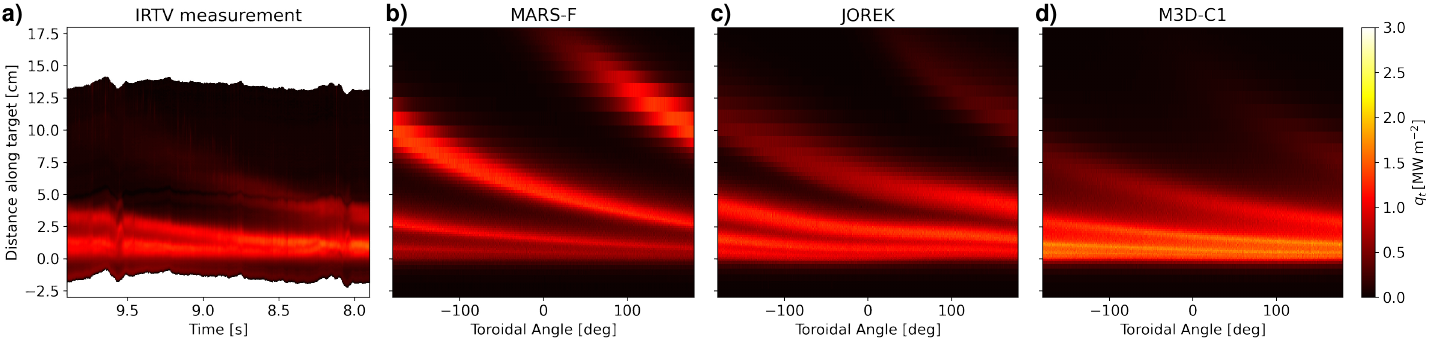}
\caption{The same \nsepx and \PSOL as in figure \ref{fig:heat_load_benchmark_nsepx05_P15}, but with reduced cross-field transport of $D = 0.15 \, \meter^2 \, \second^{-1}$ and $\chi = 0.5 \, \meter^2 \, \second^{-1}$.}
\label{fig:heat_load_benchmark_Dlow}
\end{center}
\end{figure}

For a quick exploration, simulations are repeated with a reduced power input of $\PSOL = 1.5 \, \mega\watt$ and results are shown in figure \ref{fig:heat_load_benchmark_nsepx05_P15}.
As should be expected, the heat load pattern itself does not change when \PSOL is reduced.
However, peak values are more consistent with the IRTV measurement for the lower \PSOL.
Nevertheless, even the small magnetic footprint of the M3D-C1 plasma response produces a somewhat broader pattern with lower contrast than observed by IRTV.
This may be caused by overestimating the impact of anomalous cross-field transport.
Figure \ref{fig:heat_load_benchmark_Dlow} shows the simulation results with the same \nsepx and \PSOL as in figure \ref{fig:heat_load_benchmark_nsepx05_P15}, but at half the level of cross-field transport with $D = 0.15 \, \meter^2 \, \second^{-1}$ and $\chi = 0.5 \, \meter^2 \, \second^{-1}$.
It can be seen that the patterns become sharper as less power is crossing into the traditional SOL outside the separatrix.
However, this implies that the peak values rise again - above the observed value for the M3D-C1 case.
This emphasizes the suspicion that the initial assumption of $\PSOL = 3.0 \, \mega\watt$ is too high, or that substantially more radiative power losses occur in the SOL (it later turned out that the signal for NBI2 was misrepresented such that the total heating power was in fact $\Pheat = 3.0 \, \mega\watt$, which implies that \PSOL is indeed lower after taking into account core radiation).

\begin{figure}
\begin{center}
\includegraphics[width=160mm]{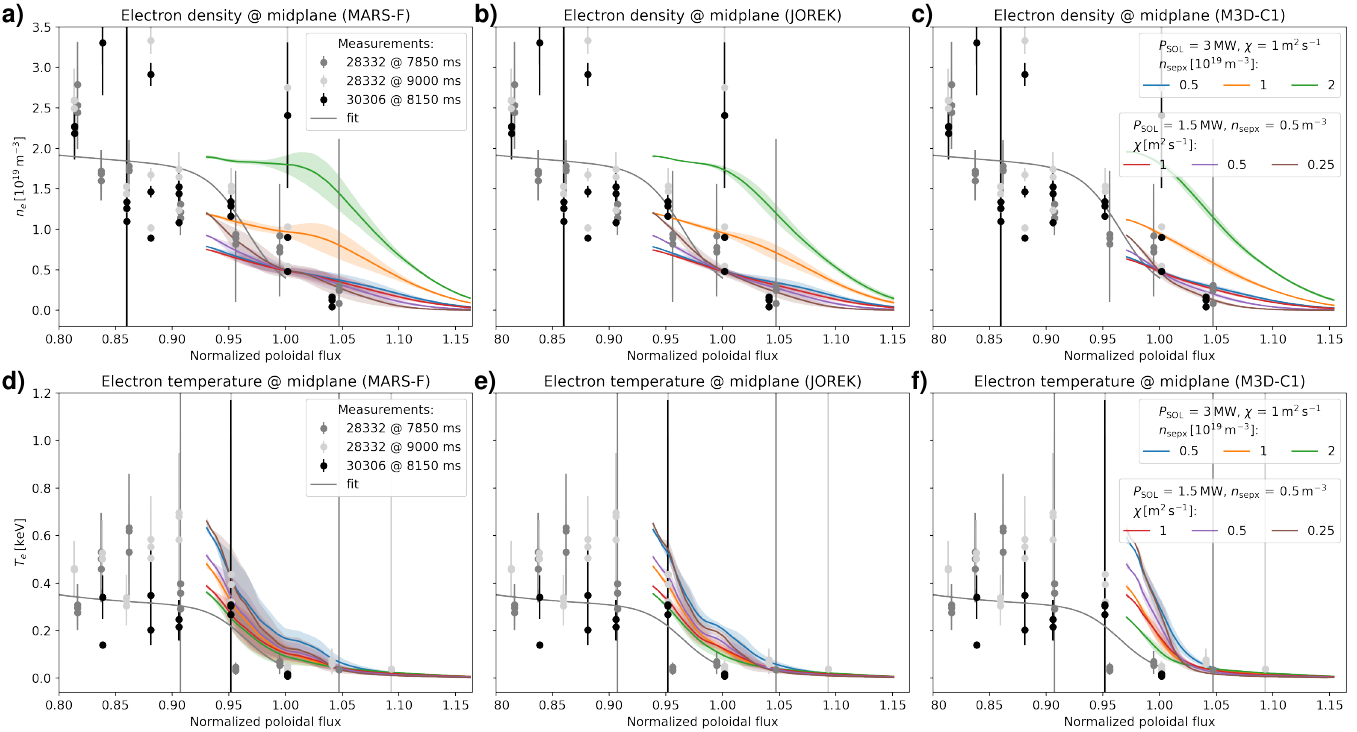}
\caption{(a-c) Electron density and (d-f) electron temperature midplane profiles per plasma response model for different simulation parameters. Thomson data for this discharge (black) and for an alternative discharge (gray) that was used as input for the equilibrium reconstruction are shown as symbols. The input equilibrium profile is shown as gray line.}
\label{fig:midplane_profiles}
\end{center}
\end{figure}

\begin{figure}
\begin{center}
\includegraphics[width=160mm]{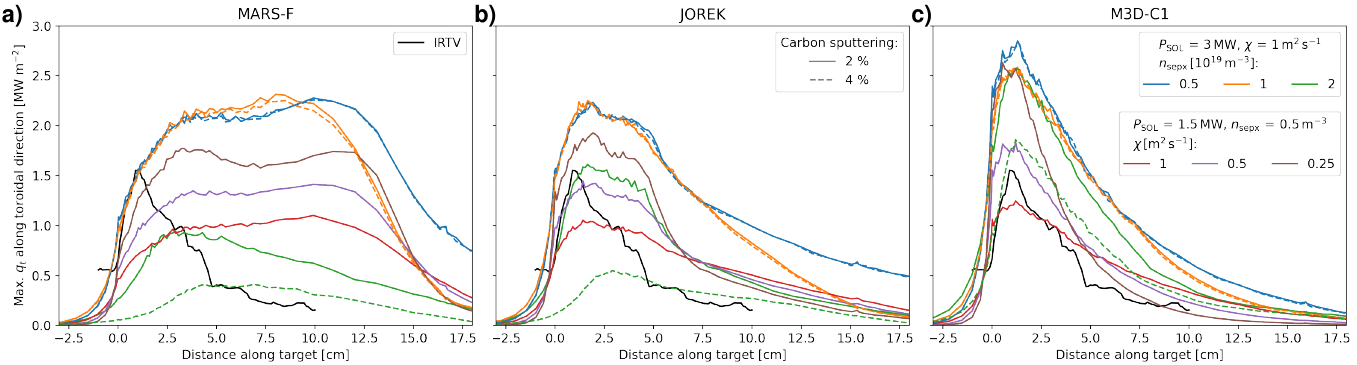}
\caption{Profiles along lobes by taking the maximum $q_t$ along the toroidal direction while moving along the target: for different densities at fixed $\PSOL \, = \, 3 \, \mega\watt$ and $\chi \, = \, 1 \, \meter^2 \, \second^{-1}$, and for different cross-field transport at fixed $\PSOL \, = \, 1.5 \, \mega\watt$ and $\nsepx \, = \, 0.5 \cdot 10^{19} \, \meter^{-3}$.
The dashed line lines show simulations with $4 \, \%$ carbon sputtering instead of $2 \, \%$ with otherwise identical parameters.
The black line shows the IRTV data for reference.}
\label{fig:heat_load_profiles}
\end{center}
\end{figure}

As mentioned above, a possible reason for underestimating the radiation is a mismatch in plasma conditions.
Up to now, $\nsepx = 0.5 \cdot 10^{19} \, \meter^{-3}$ has been used as control parameter in the simulations.
This is consistent with the fitted $n_e$ profile that was used as input for the equilibrium reconstruction (gray line in figure \ref{fig:midplane_profiles}).
However, given the uncertainty in the measurements, higher values of \nsepx are plausible.
Additional simulations are conducted with $\nsepx = 1 \cdot 10^{19} \, \meter^{-3}$ and $2 \cdot 10^{19} \, \meter^{-3}$.
Results are shown in figures \ref{fig:midplane_profiles} and \ref{fig:heat_load_profiles} in comparison with the previous simulations discussed above.
In particular, the simulated upstream temperature profiles in figure \ref{fig:midplane_profiles} (d-f) provide another indication for $\PSOL = 3.0 \, \mega\watt$ being too high.
But they also show that it becomes more difficult to match upstream conditions at lower cross-field transport that would be necessary to match the high contrast of the observed heat load striation pattern.

In figure \ref{fig:heat_load_profiles}, heat load profiles along the lobes are generated by taking the maximum $q_t$ along the toroidal direction while moving along the target.
It can be seen that increasing \nsepx from $0.5 \cdot 10^{19} \, \meter^{-3}$ (blue) to $1 \cdot 10^{19} \, \meter^{-3}$ (orange) does not significantly affect the peak heat load inside the lobes (although the downstream densities and temperatures change between these simulations).
The plasma remains attached, and power losses from impurity radiation are still negligible at $\nsepx = 1 \cdot 10^{19} \, \meter^{-3}$.
However, increasing \nsepx further to $2 \cdot 10^{19} \, \meter^{-3}$ (green) shows a reduction of the peak heat load for the JOREK and MARS-F cases - but not for the M3D-C1 case.
Impurity radiation becomes a contributing factor at this higher density ($36 \, \%$ and $44 \, \%$ of \PSOL for the JOREK and MARS-F cases, respectively) - although not as much for the M3D-C1 case ($19 \, \%$).
The peak values for the JOREK and MARS-F cases are in better agreement with the IRTV measurement, except the additional power losses are spread more or less evenly such that the resulting profile shape is still inconsistent with the measurement.

Higher downstream densities may be necessary to drive intrinsic impurity radiation for the M3D-C1 case in order to bring down the peak heat load.
However, increasing \nsepx further would be inconsistent with upstream density measurements.
But - as suggested above - another reason for the mismatch could be a higher net erosion rate.
This is because the material conditions (temperature, surface roughness, ...) play an important role for the net chemical sputtering rate.
To cover this uncertainty, simulation results where \csput has been increased from 2 to $4 \, \%$ are shown as dashed profiles in figure \ref{fig:heat_load_profiles} for comparison.
While no significant impact is found for the two lower \nsepx cases (blue, yellow), the resulting impurity radiation becomes the dominant power loss channel for the highest \nsepx (green) - with heat loads below what is seen by IRTV for the JOREK and MARS-F cases.
Finally, for the M3D-C1 case, a better agreement with measurements is found with the additional power losses at $\csput = 4 \, \%$ and $\nsepx = 2 \cdot 10^{19} \, \meter^{-3}$.
While the upstream profiles in figure \ref{fig:midplane_profiles} (a-c) suggest that $\nsepx = 2 \cdot 10^{19} \, \meter^{-3}$ may be a bit too high, it should be possible to achieve similar agreement with a more moderate \nsepx together with lower \PSOL and cross-field transport.

Finally, an additional complication is the equilibrium scenario itself: at $B_t = 1.79 \, \tesla$ and $I_p = 0.51 \, \mega\ampere$, the grazing angle of field lines on the outer lower target is just below $1 \, \deg$.
This poses a huge challenge for measurements where small misalignments of surface elements can easily account for a factor of 2 differences in heat loads.
Furthermore, the heat load striations are measured at one toroidal location by rotating the perturbation field - and thus the helical lobes of the perturbed separatrix.
A potential surface misalignment at that location introduces a systematic error for the entire striation pattern and the resulting integral value.
Therefore, also the global power balance (based on the integral value on the target surface) is left with large uncertainty and cannot be used to constrain radiative losses.


\section{Conclusions}

Modeling of the magnetic footprint for an ELM suppressed H-mode plasma at KSTAR shows that the extent of the helical striations varies from $2 \, \centi\meter$ to $14 \, \centi\meter$ depending on the plasma response model.
This uncertainty is reflected in EMC3-EIRENE simulations of the resulting heat loads: it is found
that either the peak value or the extent of the striations appear to be overestimated compared to IRTV measurements.
This suggests that the initial power into the scrape-off layer has been overestimated, and better agreement is indeed found either for lower input power and lower cross-field transport, or for higher upstream density and radiative power losses.
Future experiments with better data quality may be helpful for narrowing down inputs for the separatrix density and anomalous cross-field transport.
Coupling of the plasma boundary model to a core model would be useful for a more accurate net power input required as boundary condition in the former, in particular after the upgrade to the new tungsten divertor.
However, this must in turn include an accurate model for the surface state which determines the balance between erosion and deposition.

Although the exact reason for the mismatch between plasma response models remains unclear, several factors can contribute to these discrepancies.
For example, the GPEC and MARS-F codes utilize an equilibrium truncated inside the separatrix or smoothed near the X-point to facilitate a magnetic coordinate representation.
In contrast, the JOREK and M3D-C1 simulations extend the domain beyond the separatrix without such limitation.
This difference can lead to different field responses in the SOL, while the field response inside the separatrix remains relatively similar across models.
Even though the best match for the divertor heat loads is achieved with M3D-C1 here, it must be noted that this based on a linear, single-fluid calculation.
This motivates further benchmarking of plasma response models - including a sensitivity analysis with respect to the equilibrium profiles provided as input in linear and non-linear plasma response calculations.
One particular aspect for the latter is a equilibrium profile degradation effect by RMPs which can significantly modify the plasma response.
The difference between the JOREK calculation here and the IRTV measurement can be related to the RMP boundary constraint used here where the perturbed magnetic flux is fixed on the computational boundary.
An alternative is to use JOREK-STARWALL for a self-consistent free boundary approach which couples the plasma to the vacuum region and to conducting structures.
This is, however, left as future work.


\appendix
\section*{Acknowledgements}

This work was supported by the U.S. Department of Energy under Award No. DE-SC0020357.
The authors like to thank M. Hoelzl and the JOREK team for constructive discussions about plasma response modeling.
\section{Divertor footprints with MARS-F} \label{sec:marsf_boundary_scan}

\begin{figure}
\begin{center}
\includegraphics[width=160mm]{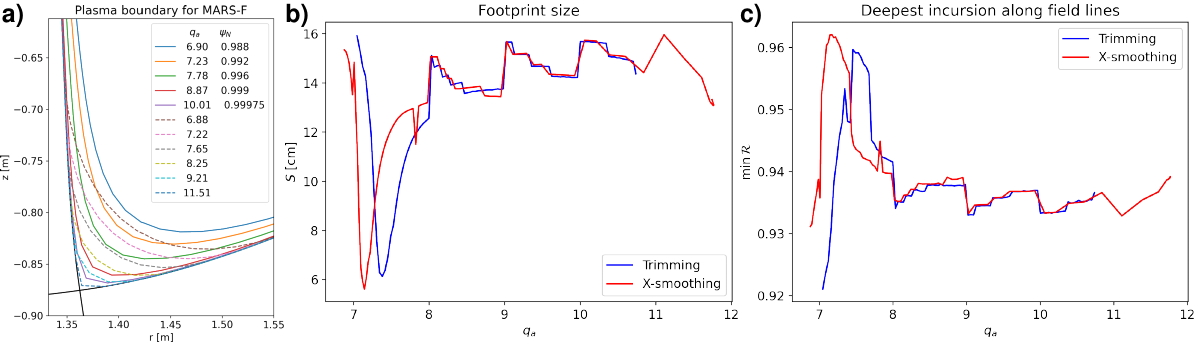}
\caption{(a) Plasma boundary contours without X-point for MARS-F: flux surfaces at selected $\psiN$ values just inside the separatrix (solid lines: Trimming), and smoothed contours by replacing parts of the separatrix with a cubic B\'ezier curve (dashed lines: X-smoothing).
(b) Resulting magnetic footprint size and (c) deepest incursion of field lines from the divertor target.}
\label{fig:boundary_scan}
\end{center}
\end{figure}

MARS-F solves the perturbed single fluid resistive MHD equations in magnetic coordinates for the plasma, together with equations for the vacuum.
The equilibrium input is mapped to the MARS-F coordinate system using the CHEASE code \cite{Lutjens1996}, a fixed boundary Grad-Shafranov equilibrium solver.
In the following we compare two methods for removing the X-point and the associated singularity of the safety factor $q$: 1) shaving of the $q$-profile near the plasma boundary by cutting away a small portion of flux surfaces, and 2) smoothing of the plasma boundary near the X-point.
Figure \ref{fig:boundary_scan} (a) shows a variation of plasma boundary contours with the resulting boundary safety-factor values \qboundary for both methods.
Similar studies have been conducted with a focus on selected field components \cite{Ryan2015, Yang2019}.
Here, on the other hand, we evaluate the impact on the resulting magnetic footprint on the outer divertor target.

For field line tracing in FLARE, cylindrical coordinates are first mapped to the MARS-F magnetic coordinates, and then the latter are used to interpolated the plasma response at any given point.
This is then superposed onto the equilibrium field in poloidal divertor geometry.
Figure \ref{fig:boundary_scan} (b-c) shows that the footprint size and radial excursion of field lines can be sensitive to the choice of the plasma boundary contour if too much smoothing or trimming is applied (below $\psiN \approx 0.996$ or $\qboundary \approx 8$).
On the other hand, less variation is found for $\qboundary > 8$, and calculations tend to be more robust each time \qboundary approaches the next integer value.
The same radial and poloidal resolution is used for all calculations here, and this imposes a limit on \qboundary values due to numerical accuracy.


\section*{References}


\end{document}